\documentclass[prc,twocolumn,showpacs,amsmath,amssymb]{revtex4}
\usepackage{times}
\usepackage{graphicx}
\usepackage{dcolumn}
\usepackage{bm}
\usepackage{amstext}

\begin{document}

\title{Superdeformed structures and low $\Omega$ parity doublet
in Ne$-$S nuclei near neutron drip-line }

\author{Shailesh K. Singh, S.K. Patra and C.R. Praharaj}
\affiliation{Institute of Physics, Bhubaneswar-751 005, 
India} 

\date{\today}
\begin{abstract}

The structure of Ne, Na, Mg, Al, Si, P and S nuclei near the neutron drip-line
region is investigated in the frame-work of relativistic mean field 
theory and non-relativistic Skyrme Hartree-Fock formalism. 
The recently discovered nuclei $^{40}$Mg and $^{42}$Al, which 
are beyond the drip-line predicted by various mass formulae are 
located within these models.
We find many 
largely deformed neutron-rich nuclei, whose structures are analyzed. 
From the structure anatomy, we find that at large deformation, 
low $\Omega$ orbits of opposite parities (e.g. 
$\frac{1}{2}^+$ and $\frac{1}{2}^-$) occur close to each other in energy.

\end{abstract}
\maketitle


\section{Introduction}

The structure of light nuclei near the neutron drip-line is one of the
interesting topic 
for a good number of exotic phenomena. Nuclei in this region
are quite different in collectivity and clustering features than the stable
counterpart in the nuclear chart.  For example, the neutron magic property 
is lost for N = 8 in $^{12}$Be \cite{navin00} and N = 20 in $^{32}$Mg
\cite{motobayashi95}. The unexpectedly large reaction cross-section 
for $^{22}$C gives the indication of neutron halo structure \cite{tanaka10}.
The discovery of large collectivity of $^{34}$Mg by Iwasaki et al.
\cite{iwasaki01} is
another example of such exotic properties. The deformed structures, core
excitation and the location of drip-line for Mg and neighboring nuclei 
are few of the interesting properties of investigation.
In this context, the discovery of $^{40}$Mg and $^{42}$Al, once predicted
to be nuclei beyond the drip-line by various mass formulae 
\cite{baumann07,moller95} 
show to need  the modification of the mass models.

On the other hand, the appearance of N = 16 as magic number in $^{24}$O and 
the existence of neutron halo in $^{11}$Li are established observations
\cite{ozawa00}. However, the proposed proton \cite{mina} ($^{8}$B) 
and neutron \cite{tan96,nakamura11} halo 
($^{14}$Be, $^{17}$B, $^{31}$Ne) in the exotic nuclei
are currently under investigations. In addition to these, the cluster 
structure of entire the light mass nuclei and the
skin formation in neutron-drip isotopes motivate us 
for the study of light mass drip-line nuclei. In this paper, our aim is
to study the neutron drip-line for Ne$-$S isotopic chain in the frame-work
of a relativistic mean field (RMF) and nonrelativistic Skyrme Hartree-Fock (SHF)
formalisms and analyzed the features of large quadrupole deformation of
these isotopes.

The paper is organized as follows: The RMF and SHF formalisms are described 
briefly in Section II. The results obtained from our calculations are 
discussed in 
Section III. Finally, summary and concluding remarks are given in Section IV.

\section{The Formalism}
The mean field methods like SHF and RMF have been widely used in the 
study of binding
energies, root mean square radii, quadrupole deformation and other bulk
properties of nuclei \cite{vautherin72,reinhard89}.
In general, one can say that although the older parametrizations of 
SHF and RMF have some
limitation to predict the experimental observables, the recent forces
are good enough to reproduce the bulk properties not only 
near the $\beta-$stability line but also far away from it.
Here, we use these two successful models
\cite{vautherin72,reinhard89,ring90,bender03,lunney03,vretenar05,meng06,
niksic11,paar07,erler11,takashi12,cha97,stone03,stone07,rei95,cha98,sero86} 
to learn about the properties of drip-line nuclei Ne$-$S.

\subsection{The Skyrme Hartree-Fock (SHF) Method}
The general form of the Skyrme effective interaction used in the
mean-field model can be expressed as a Hamiltonian density $\cal H$
\cite{cha97,bender03,lunney03,stone03,stone07,erler11,takashi12}. 
This $\cal H$ is written expressed as a function of some 
empirical parameters given as:
\begin{equation}
{\mathcal H}={\mathcal K}+{\mathcal H}_0+{\mathcal H}_3+ {\mathcal H}_{eff}
+\cdots, \label{eq:1}
\end{equation}
where ${\cal K}$ is the kinetic energy term, ${\cal H}_0$ the zero range, 
${\cal H}_3$ the density dependent and ${\cal H}_{eff}$ the effective-mass 
dependent terms, which are relevant for calculating the properties of 
nuclear matter. These are functions of 9 parameters $t_i$, $x_i$ 
($i = 0,1,2,3$) and $\eta$, and are given as
\begin{eqnarray}
{\mathcal H}_0&=&\frac{1}{4}t_0\left[(2+x_0)\rho^2 - (2x_0+1)(\rho_p^2+\rho_n^2)\right],
\label{eq:2}
\\
{\mathcal H}_3&=&\frac{1}{24}t_3\rho^\eta \left[(2+x_3)\rho^2 - (2x_3+1)(\rho_p^2+\rho_n^2)\right],
\label{eq:3}
\\
{\mathcal H}_{eff}&=&\frac{1}{8}\left[t_1(2+x_1)+t_2(2+x_2)\right]\tau \rho \nonumber \\
&&+\frac{1}{8}\left[t_2(2x_2+1)-t_1(2x_1+1)\right](\tau_p \rho_p+
\tau_n \rho_n). 
\label{eq:4}
\end{eqnarray}
The kinetic energy ${\cal K} = \frac{\hbar^2}{2M}\tau$, a form used in 
the Fermi gas model for non-interacting Fermions. The other terms, 
representing the surface contributions of a
finite nucleus with $b_4$ and $b^{\prime}_4$ as additional parameter, are
\begin{eqnarray}
{\mathcal H}_{S\rho}&=&\frac{1}{16}\left[3t_1(1+\frac{1}{2}x_1)-t_2(1+\frac{1}{2}x_2)\right](\vec{\nabla}\rho)^2 \nonumber\\
&&-\frac{1}{16}\left[3t_1(x_1+\frac{1}{2})+t_2(x_2+\frac{1}{2})\right] \nonumber\\
&&\times\left[(\vec{\nabla}\rho_n)^2+(\vec{\nabla}\rho_p)^2\right],
\label{eq:5}
\\
{\mathcal H}_{S\vec{J}}&=&-\frac{1}{2}\left[{b_4}\rho\vec{\nabla}\cdot\vec{J}+{b^{\prime}_4}(\rho_n\vec{\nabla}\cdot\vec{J_n}
+\rho_p\vec{\nabla}\cdot\vec{J_p})\right].
\label{eq:6}
\end{eqnarray}
Here, the total nucleon number density $\rho=\rho_n+\rho_p$, the kinetic energy
 density $\tau=\tau_n+\tau_p$, and the spin-orbit density 
$\vec{J}=\vec{J}_n+\vec{J}_p$. The subscripts $n$ and $p$ refer to neutron and 
proton, respectively. The nucleon mass is represented by $m$.
The $\vec{J}_q=0$, $q=n$ or $p$, for spin-saturated nuclei, i.e., for nuclei
with major oscillator shells completely filled or empty. The total binding 
energy (BE) of a nucleus is the integral of $\cal H$.

\subsection{The Relativistic Mean Field (RMF) Method}
The relativistic mean field approach is well-known and the theory is well 
documented \cite{sero86,ring90,vretenar05,meng06,paar07,niksic11}. 
Here we start with the relativistic Lagrangian
density for a nucleon-meson many-body system as:
\begin{eqnarray}
{\cal L}&=&\overline{\psi_{i}}\{i\gamma^{\mu}
\partial_{\mu}-M\}\psi_{i}
+{\frac12}\partial^{\mu}\sigma\partial_{\mu}\sigma
-{\frac12}m_{\sigma}^{2}\sigma^{2}\nonumber\\
&& -{\frac13}g_{2}\sigma^{3} -{\frac14}g_{3}\sigma^{4}
-g_{s}\overline{\psi_{i}}\psi_{i}\sigma-{\frac14}\Omega^{\mu\nu}
\Omega_{\mu\nu}\nonumber\\
&&+{\frac12}m_{w}^{2}V^{\mu}V_{\mu}
-g_{w}\overline\psi_{i}
\gamma^{\mu}\psi_{i}
V_{\mu}\nonumber\\
&&-{\frac14}\vec{B}^{\mu\nu}.\vec{B}_{\mu\nu}+{\frac12}m_{\rho}^{2}{\vec
R^{\mu}} .{\vec{R}_{\mu}}
-g_{\rho}\overline\psi_{i}\gamma^{\mu}\vec{\tau}\psi_{i}.\vec
{R^{\mu}}\nonumber\\
&&-{\frac14}F^{\mu\nu}F_{\mu\nu}-e\overline\psi_{i}
\gamma^{\mu}\frac{\left(1-\tau_{3i}\right)}{2}\psi_{i}A_{\mu} .
\end{eqnarray}
All the quantities have their usual meanings. From the relativistic
Lagrangian, we obtain the field equations for the nucleons and mesons. These
equations are solved by expanding the upper and lower components of the Dirac
spinor and the boson fields in an axially deformed harmonic oscillator basis.
The set of coupled equations is solved numerically by a self-consistent 
iteration method. The total energy of the system in RMF formalism is given by
\begin{equation}
 E_{total} = E_{part}+E_{\sigma}+E_{\omega}+E_{\rho}+E_{c}+E_{pair}+E_{c.m.},
\end{equation}
where $E_{part}$ is the sum of the single particle energies of the nucleons and
$E_{\sigma}$, $E_{\omega}$, $E_{\rho}$, $E_{c}$, $E_{pair}$, $E_{cm}$ are
the contributions of the meson fields, the Coulomb field, pairing energy
and the center-of-mass energy, respectively.

\subsection{Pairing correlation }
To take care of the pairing correlation for open shell nuclei, the
constant gap BCS-approach is used in our calculations. 
The pairing energy expression is written as:
\begin{equation}
E_{pair}=-G\left[\sum_{i>0}u_{i}v_{i}\right]^2,
\end{equation}
with $G$=pairing force constant, $v_i^2$ and $u_i^2=1-v_i^2$ are the occupation
probabilities\cite{pres82,patra93}. The variational approach with
respect to $v_i^2$ gives the BCS equation \cite{pres82}
\begin{equation}
2\epsilon_iu_iv_i-\triangle(u_i^2-v_i^2)=0,
\end{equation}
using $\triangle=G\sum_{i>0}u_{i}v_{i}$. The occupation number is defined as:
\begin{equation}
n_i=v_i^2=\frac{1}{2}\left[1-\frac{\epsilon_i-\lambda}{\sqrt{(\epsilon_i-\lambda)^2+\triangle^2}}\right].
\end{equation}
The chemical potentials $\lambda_n$ and $\lambda_p$ are determined by the 
particle numbers for neutrons and protons. The pairing energy is computed as 
$E_{pair}=-\triangle\sum_{i>0}u_{i}v_{i}$.
For a particular value of $\triangle$ and $G$, the pairing energy $E_{pair}$ 
diverges, if it is extended to an infinite configuration space. 
In fact, in all realistic calculations with finite range forces,
the contribution of states of large momenta above the Fermi surface
(for a particular nucleus) to  $\triangle$ decreases with energy.
We use a pairing window, where the equations are extended up to the level 
$|\epsilon_i-\lambda|\leq 2(41A^{-1/3})$.
The factor 2 has been determined so as to reproduce the
pairing correlation energy for neutrons in $^{118}$Sn using Gogny force
\cite{sero86,patra93,dech80}. The values of $\triangle_n$ and $\triangle_p$ are 
taken from \cite{madland}, as input in the BCS-eqaution. 

We compare the results with various simple and sophisticated pairing 
prescriptions like BCS-delta force \cite{bcs-delta} and BCS density 
dependent delta force \cite{delta}. 
These calculations have been done only for $^{20}$Ne and $^{47}$Al nucleus 
in both SkI4 and NL3 force parameter sets. We have given these results 
in Table \ref{tab:pair} along with experimental results like quadrupole 
deformation parameter $\beta_2$ \cite{raman}, total binding energy (BE) 
\cite{audi12} and root mean square charge radius ($r_{ch}$) \cite{angeli13}.
We find that, for this lighter mass region of the periodic chart, 
pairing is less important for majority of the cases. 
With pairing, the deformation becomes negligible for $^{20}$Ne and we do not 
get the experimental deformation parameter in RMF calculations. 
With no pairing, we reproduce substantially the deformation parameter in RMF 
because the 
{\it density of states} near Fermi surface for such light nuclei are small and  
not conducive to pairing \cite{ripka}. 
To understand the influence of pairing on the open shell nuclei, 
we have taken into account the experimental data, wherever available. The 
SHF(SkI4) results are used as guideline in the absence of these data.  
We realized after comparing the calculated $\beta_2$ of RMF and
SHF with experimental data that the quadrupole deformation of SHF
is closer to experiment without taking pairing correlation into account. 
For example, when we use the
$\triangle_n$ and $\triangle_p$ from the experimental binding energy of
odd-even values or from the empirical formula of Ref. \cite{madland,mott69}
to calculate $\beta_2$ for $^{20,22,24,26,28}$Ne in RMF(NL3), we find 
$\beta_2\sim 0.18, 0.35, 0.19, 0.0, 0.0$, respectively for
these isotopes agreeing with the result of Lalazissis et al \cite{lala99}. 
These $\beta_2$ strongly disagree with the measured values 
($\beta_2(expt.) = 0.723, 0.562, 0.45, 0.498, 0.50$) \cite{raman}. 
Similar effects are also seen in other considered isotopes. 
On the other hand, if we ignore pairing, then the calculated results are 
often better and these $\beta_2$ are quite close to
the experimental data. The influence of pairing is also visible in the
total binding energy. In some of the cases, even a couple
of MeV difference in total binding energy is found with and without taking
pairing correlation into account in RMF formalism. Contrary to the RMF, 
the pairing in the SHF formalism
is almost insensitive to quadrupole deformation for the considered
mass region. 
Thus, we have performed the calculations through out the paper without 
considering pairing into account.

\begin{table}
\caption{\label{tab:pair}Calculation of binding energy (BE), quadrupole 
deformation parameter $\beta_2$, root mean square of matter radius ($r_{rms}$)
 and 
charge radius ($r_{ch}$) by taking various pairing methods. We have given 
these results for both SkI4 and NL3 parameter sets with experimental data
\cite{raman,audi12,angeli13}. } 
\begin{tabular}{|cccccc|}
\hline
SkI4 \\
\hline
Nucl. &  Type of Pairing & $\beta_2$ & $r_{rms}$ &  $BE$   & $r_{ch}$ \\
\hline
$^{20}$Ne & No pairing         & 0.549    & 2.911   & 156.8 & 3.030  \\
          & BCS-delta force    & 0.548    & 2.910   & 156.8 & 3.030   \\
          & BCS-dens.dep.delta-force       &0.548 &2.910  & 156.8 & 3.030  \\
$^{47}$Al & No pairing         & 0.006     & 3.972   & 287.8 & 3.324  \\
          & BCS-delta force    & 0.007     &3.957    & 288.7 & 3.317  \\
          & BCS-dens.dep.delta-force&0.055 &3.970    & 288.0 & 3.322  \\
\hline
NL3 \\
\hline
$^{20}$Ne & No pairing           & 0.537  & 2.846   & 156.7 & 2.972  \\
          & BCS-delta           & 0.036  & 2.920   & 154.9 & 3.055  \\
$^{47}$Al & No pairing          & 0.090  & 3.832   & 294.6 & 3.246   \\
          & BCS-delta           & 0.081  & 3.834   & 294.8 & 3.246  \\
\hline
Exp. Results  \\
\hline
$^{20}$Ne &                    & 0.728      &         & 160.6 & 3.005  \\
$^{47}$Al &                    &            &         & ---     &         \\
\hline
\end{tabular}
\end{table}
\subsection{Pauli Blocking and Harmonic oscillator basis }
For even-even nucleus $\pm m$ orbits are pairwise occupied and the mean 
field has time reversal symmetry. 
But in the case of odd nucleon the time reversal symmetry is broken. 
To take care of the odd nucleon, we employ the blocking method \cite{patra01}.
We put the last nucleon in one of the conjugate 
states $\pm$m and keeping other state empty. In this way we follow the 
time reversal symmetry for odd-even and odd-odd nuclei. We repeat this 
calculation by putting the odd nucleon in all the near by state of the 
conjugate level to determine the maximum binding energy of the ground 
state \cite{patra01, patra91}.

\begin{figure}
\caption{\label{fig:basis} The change in binding energy BE, root mean square 
matter radius ($r_{rms}$), quadrupole deformation parameter $\beta_2$
with Fermionic $N_{F}$ and bosonic $N_{B}$ harmonic oscillator basis 
for some selected nuclei. }
\includegraphics[scale=0.35]{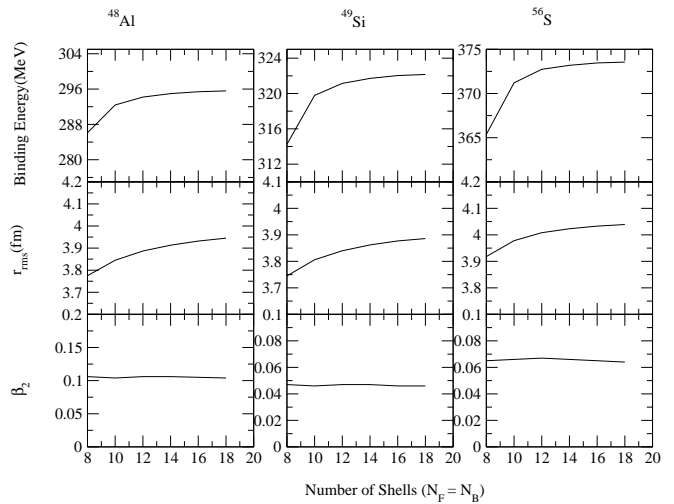}
\end{figure}

In our present calculations the nuclei are treated as axial-symmetrically 
deformed, with $z$-axis as the symmetric axis. Spherical symmetry is no longer 
present in general and therefore $j$ is not a good quantum number any more.
Because of axial symmetry, each orbit is denoted by the quantum number $m$ 
of $J_z$ and is a superposition of $|jm>$ states with various $j$ values.
The densities are invariant with respect to a rotation around the symmetry axis. 
For numerical calculations, 
the wavefunctions are expanded in a deformed harmonic oscillator
potential basis and solved self-consistently in an iteration method.
The major oscillator quanta for Fermion $N_F$ and bosons $N_B$
are taken as $N_{max}$ = 12. The convergence of our 
numerical results are tested in Fig. \ref{fig:basis} for BE, matter radius 
 $r_{rms}$ and quadrupole deformation parameter for some selected nuclei 
like $^{48}$Al, $^{49}$Si and $^{56}$S. Here, the results are estimated from 
$N_F = N_B = 8$ to $N_F = N_B = 18$, which are shown in Fig. \ref{fig:basis}.
From this analysis, we observed that the $\beta_2$ values almost identical
with the variation of oscillator quanta.
However, the rms radii and binding energy vary till $N_F = N_B = 12$,
beyond which the results are unchanged. 
It is well known that harmonic oscillator basis is not suitable in dripline 
nuclei due to the asymptotic behavior of the density distribution. 
To resolve this problem , efforts have been made for solving the 
equations in coordinate space \cite{doba84,meng96,del01}. 
Some other kinds of basis like transformed harmonic oscillator basis 
\cite{stoi98}, Gaussian expansion method \cite{nakada02} and  
Woods-Saxon basis \cite{shan03,shan10} are also available in literature.  
The inclusion of sufficiently large harmonic oscillator  model space 
gives reasonably 
convergent results. This type of prescription is already done in Ref. 
\cite{patra05}. However, to fully include continuum effects 
more work has to be done (by use of basis of finite potentials and 
inclusion of correlation effects in Hartree-Bogoliubov scheme \cite{hfb}).


\subsection{Ground state properties from the SHF and RMF models}
Certainly for light mass nuclei,
the correction of centre of mass motion can not be ignored and it should
be done self-consistently. That means, in the evaluation of centre-of-mass
energy, one should evaluate $E_{CM}=\frac{<F|P^2|F>}{2M}$ using
$|F>=|F>_{RMF}$ wavefunction. In this case, one has to calculate the matrix
elements directly. However, this procedure is more involved  and in the
present calculations we have subtracted the spurious centre-of-mass motion
using the Elliott-Skyrme approximation and the approximate analytical expression
is written as $E_{CM}=\frac{3}{4}.41A^{-1/3}$ MeV (harmonic oscillator
approximation), where A is mass number \cite{elliott55,negele70,toki94} and 
expect that the two results 
should not differ drastically. 
The quadrupole moment deformation parameter $\beta_2$ is evaluated from the 
resulting proton and neutron quadrupole moments through:
\begin{equation} 
Q=Q_n+Q_p=\sqrt{\frac{16\pi}5} \left(\frac3{4\pi} AR^2\beta_2\right),
\end{equation}
where R = 1.2$A^{1/3}$.
The root mean square radii of protons and matter distribution are 
defined as 
$\langle r_p^2\rangle={1\over{Z}}\int\rho_p(r_{\perp},z) r^2d\tau$,
and  
$\langle r_{rms}^2\rangle={1\over{A}}\int\rho(r_{\perp},z) r^2d\tau$;
respectively, where $Z$ is the proton number and $\rho_p(r_{\perp},z)$ 
is the deformed proton and $\rho(r_{\perp},z)$ is the total nucleon density 
distribution. The proton and charge rms radius is connected through 
the relation $r_{ch}$ = $\sqrt{r_p^2 + 0.64}$ \cite{patra91}.

We use the well known NL3 parameter set \cite{lala97} for the RMF formalism.
This set not only reproduces the 
properties of stable nuclei but also well predicts for those far from the
$\beta$-stability valley. Also, the isoscalar monopole energy agrees 
excellently with the experimental values for different regions of the 
Periodic Table. The measured
superdeformed minimum in $^{194}$Hg is 6.02 MeV above the ground state,
whereas in RMF calculation with NL3 set, this number is 5.99 MeV \cite{lala97}.

For SHF model, 
we use the Skyrme SkI4 set with $b_4\ne b^{\prime}_4$ \cite{rei95}. 
This parameter set is designed for considerations of proper spin-orbit 
interaction in finite nuclei,
related to the isotope shifts in Pb region and is better suited for the
study of exotic nuclei. Several more recent Skyrme parameters such as
SLy1-10, SkX, SkI5 and SkI6 are obtained by fitting the Hartree-Fock (HF) 
results with experimental data for nuclei starting from the valley 
of stability to neutron and proton drip-lines \cite{cha97,rei95,cha98,brown98}.

\section{Results and Discussions}
The binding energy BE, rms charge radius $r_{ch}$ and quadrupole deformation 
parameter $\beta_2$ of the isotopes of Ne, Na, Mg, Al, Si, P and S
are calculated near the drip-line region.
For this, both the relativistic and non-relativistic models are used. 
%
\subsection{Binding energy and neutron drip-line}
The ground state binding energy (BE) for
Ne, Na, Mg, Al, Si, P  and S isotopes are selected by comparing the 
binding energy obtained from the prolate, oblate and spherical solutions
for a particular nucleus. For a given nucleus, the
maximum binding energy corresponds to the ground state and other
solutions are obtained as various excited intrinsic states.
In Table \ref{tab:table1}, the ground state binding energy for the heaviest
known isotopes for the discussed nuclei are compared with the experimental
data \cite{audi12}. The binding energy for $^{31}$Ne is 216.0 MeV with
RMF (NL3) and these are 213.2 and 211.4 MeV in SHF(SkI4) and
experiment, respectively. Similarly, these results for $^{45}$S respectively 
are 353.4, 350.4 and 354.7 MeV in RMF, SHF and experiment. Analyzing
the data of Table \ref{tab:table1}, generally one finds that BE of RMF 
is slightly over 
estimated and SHF is underestimated than the experimental values. However,
the overall agreement of the calculated energies are within an acceptable
range with the experimental data.  

\begin{table}
\hspace{2.0cm}
\caption{\label{tab:table1}The calculated ground state binding energy
obtained from SHF and RMF theory are compared with the experimentally
known heaviest isotope for Ne, Na, Mg, Al, Si, P and S \cite{audi12}.}
\begin{tabular}{|c|c|c|c|c|c|c|c|c|}
\hline
Nucl. &RMF&SHF&Expt.&Nucleus&RMF&SHF&Expt.\\
\hline
$^{31}$Ne& 216.0 & 213.2 & 211.4 & $^{32}$Na & 234.5 & 233.4 & 230.9 \\
$^{36}$Mg& 263.9 & 260.2 & 260.8 & $^{38}$Al & 283.5 & 281.4 & 280.3 \\
$^{41}$Si& 310.1 & 307.2 & 307.9 & $^{43}$P  & 331.7 & 329.0 & 330.7  \\
$^{45}$S & 353.4 & 350.4 & 354.7 &           &       &       &         \\
\hline
\end{tabular}
\end{table}
We have listed the neutron drip-lines in Table \ref{tab:table2}, which are obtained from the
ground state binding energy for neutron rich Ne, Na, Mg, Al, Si, P and S nuclei.
The drip-line is determined by setting the condition that the minimum 
value of two neutron separation energy $S_{2n} = BE(N,Z)-BE(N-2,Z)\geq 0$. 
The nuclei with the largest neutron numbers so far  experimentally
detected in an isotopic chain along with the extrapolated data are also 
displayed in the last column of Table \ref{tab:table2}.
The numbers given in the parenthesis are the experimentally extrapolated
values \cite{audi12}. To get a qualitative understanding of the prediction
of neutron drip-line, we have compared our results with the infinite
nuclear matter (INM) \cite{sn} and finite range droplet model (FRDM)
\cite{frdm} mass estimation. The RMF and SHF drip-lines coincide
with each other for Ne, Mg, Al and S. In case of Na and Si the RMF drip nuclei
are found to be 3 and 6 unit heavier than the SHF prediction. 
The INM prediction of drip nuclei are always in the heavier side than FRDM.
From Table \ref{tab:table2}, we find that the experimental effort has almost
reached to the INM and FRDM prediction of drip nuclei for lighter mass region.

\begin{table}
\hspace{1.9cm}
\caption{\label{tab:table2}The predicted mass number of neutron drip-line for Ne,
Na, Mg, Al, Si, P and S nucleus in RMF (NL3) and SHF (SKI4) parameter sets are
compared with infinite nuclear matter (INM) mass model \cite{sn},
finite range droplet model (FRDM) \cite{frdm} and the nuclei with 
the largest neutron numbers so far experimentally detected \cite{audi12} 
along with the number shown in parenthesis are the
experimentally extrapolated values. }
\begin{tabular}{|c|c|c|c|c|c|c|c|c|}
\hline
Nucl. &RMF&SHF&INM&FRDM&Expt.\\
\hline
Ne & 34 & 34 & 34 & 33 & 31 (34) \\
Na & 40 & 37 & 37 & 36 & 32 (37)  \\
Mg & 40 & 40 & 39 & 40 & 36 (40) \\
Al & 48 & 48 & 42 & 42 & 38 (43)  \\
Si & 54 & 48 & 45 & 43 & 41 (45) \\
P  & 54 & 55 & 49 & 48 & 43 (47) \\
S  & 55 & 55 & 51 & 51 & 45 (49)  \\
\hline
\end{tabular}
\end{table}

\begin{table}
\caption{\label{tab:table3}The calculated value of charge radius ($r_{ch}$),
quadrupole moment deformation parameter $\beta_2$ and binding energy 
(BE) for Ne, Na and Mg  nuclei in RMF (NL3) and SHF (SkI4) formalisms. 
We compare our results with experimental $\beta_2$ \cite{raman}, 
ground state binding energy BE (MeV) \cite{audi12} and the charge radius 
$r_{ch} (fm)$ \cite{angeli13}.
}
\renewcommand{\arraystretch}{0.8}
\begin{tabular}{|c|c|c|c|c|c|c|c|c|c|c|c|}
\hline
Nucl. & \multicolumn{3}{|c|}{RMF (NL3)} & \multicolumn{3}{|c|}{SHF (SkI4)}
     &  \multicolumn{3}{|c|}{Exp.} \\
\cline{2-10} 
          &$r_{ch}$&$\beta_2$&BE&$r_{ch}$&$\beta_2$&BE& $r_{ch}$&$\beta_2$&BE\\
\hline
$^{20}$Ne & 2.970 & 0.535 & 156.7 &3.030 &0.550	 &156.8	&3.006	&0.727	&160.6	\\
$^{21}$Ne & 2.953 & 0.516 & 165.9 &3.012 &0.529	 &166.8	&2.970	&		&167.4	\\
$^{22}$Ne & 2.940 & 0.502 & 175.7 &3.010 &0.520	 &175.8	&2.953	&0.562 	&177.8	\\
$^{23}$Ne & 2.913 & 0.386 & 181.8 &2.975 &0.382	 &182.2	&2.910	&		&183.0	\\
$^{24}$Ne & 2.881 & -0.259& 189.1 &2.950 &-0.250 &188.5	&2.901	&0.45   &191.8	\\
$^{25}$Ne & 2.907 & 0.272 & 194.2 &2.948 &0.170	 &194.2	&2.932	&		&196.0	\\
$^{26}$Ne & 2.926 & 0.277 & 199.9 &2.950 &0.120	 &199.4	&2.925	&0.498	&201.6	\\
$^{27}$Ne & 2.945 & 0.247 & 203.9 &2.987 &0.159	 &203.2	&	&		&203.1	\\
$^{28}$Ne & 2.965 & 0.225 & 208.2 &3.010 &0.160	 &206.5	&2.964	&0.50   &206.9	\\
$^{29}$Ne & 2.981 & 0.161 & 211.2 &3.027 &0.010	 &210.1	&	&		&207.8	\\
$^{30}$Ne & 2.998 & 0.100 & 215.0 &3.050 &0.000	 &213.7	&	&		&211.3	\\
$^{31}$Ne & 3.031 & 0.244 & 216.0 &3.057 &0.225	 &213.2	&	&		&211.4	\\
$^{32}$Ne & 3.071 & 0.373 & 218.6 &3.100 &0.380	 &213.1	&	&		&	\\
$^{33}$Ne & 3.095 & 0.424 & 219.5 &3.148 &0.429	 &213.5	&	&		&	\\
$^{34}$Ne & 3.119 & 0.473 & 220.9 &3.180 &0.490	 &213.5	&	&		&	\\
$^{24}$Na & 2.964 & 0.379 & 192.3& 3.042 &0.411	 &194.0	&2.974	&		&193.5	\\
$^{25}$Na & 2.937 & 0.273 & 200.6& 3.024 &0.314	 &201.4	&2.977	&		&202.5	\\
$^{26}$Na & 2.965 & 0.295 & 207.1& 3.027 &0.274	 &208.4	&2.993	&		&208.1	\\
$^{27}$Na & 2.993 & 0.323 & 214.2& 3.043 &0.282	 &214.9	&3.014	&		&214.8	\\
$^{28}$Na & 2.993 & 0.272 & 219.0& 3.058 &0.234	 &219.7	&3.040	&		&218.4	\\
$^{29}$Na & 3.004 & 0.232 & 224.3& 3.072 &0.194	 &224.3	&3.092	&		&222.8	\\
$^{30}$Na & 3.031 & 0.169 & 228.1& 3.079 &0.030	 &228.6	&3.118	&		&225.1	\\
$^{31}$Na & 3.047 & 0.108 & 232.7& 3.103 &0.000	 &233.5	&3.170	&		&229.3	\\
$^{32}$Na & 3.077 & 0.237 & 234.5& 3.121 &0.187	 &233.4	&	&		&230.9	\\
$^{33}$Na & 3.113 & 0.356 & 237.9& 3.172 &0.352	 &234.9	&	&		&	\\
$^{34}$Na & 3.137 & 0.404 & 239.8& 3.198 &0.407	 &236.2	&	&		&	\\
$^{35}$Na & 3.161 & 0.450 & 242.3& 3.224 &0.457	 &237.4	&	&		&	\\
$^{36}$Na & 3.175 & 0.481 & 242.5& 3.235 &0.501	 &237.5	&	&		&	\\
$^{37}$Na & 3.190 & 0.512 & 243.1& 3.251 &0.541	 &237.6	&	&		&	\\
$^{38}$Na & 3.199 & 0.491 & 243.4&	 &	 &	&	&		&	\\
$^{39}$Na & 3.209 & 0.472 & 244.1& 	 &	 &	&	&		&	\\
$^{40}$Na & 3.228 & 0.477 & 243.4& 	 &	 &	&	&		&	\\
$^{24}$Mg & 3.043 & 0.487 & 194.3& 3.130 &0.520	 &195.2	&3.057	&0.605	&198.3	\\
$^{25}$Mg & 3.009 & 0.376 & 202.9& 3.103 &0.432	 &204.3	&3.028	&		&205.6	\\
$^{26}$Mg & 2.978 & 0.273 & 212.5& 3.080 &-0.300 &213.2	&3.034	&0.482	&216.7	\\
$^{27}$Mg & 3.015 & 0.310 & 220.2& 3.096 &0.339	 &221.5	&	&		&223.1	\\
$^{28}$Mg & 3.048 & 0.345 & 228.7& 3.110 &0.340	 &229.0	&	&0.491	&231.6	\\
$^{29}$Mg & 3.055 & 0.289 & 234.3& 3.118 &0.283	 &235.0	&	&		&235.3	\\
$^{30}$Mg & 3.062 & 0.241 & 240.5& 3.120 &-0.180 &240.5	&	&0.431	&241.6	\\
$^{31}$Mg & 3.075 & 0.179 & 245.1& 3.123 &0.030	 &246.1	&	&		&243.9	\\
$^{32}$Mg & 3.090 & 0.119 & 250.5& 3.150 &0.000	 &252.0	&	&0.473	&249.7	\\
$^{33}$Mg & 3.117 & 0.233 & 253.1& 3.165 &0.155	 &253.0	&	&		&252.0	\\
$^{34}$Mg & 3.150 & 0.343 & 257.3& 3.210 &0.330	 &255.1	&	&		&256.7	\\
$^{35}$Mg & 3.173 & 0.388 & 260.5& 3.239 &0.393	 &257.8	&	&		&257.5	\\
$^{36}$Mg & 3.198 & 0.432 & 263.9& 3.265 &0.440	 &260.2	&	&		&260.8	\\
$^{37}$Mg & 3.212 & 0.462 & 264.9& 3.279 &0.469	 &261.0	&	&		&	\\
$^{38}$Mg & 3.227 & 0.492 & 266.3& 3.295 &0.490	 &261.6	&	&		&	\\
$^{39}$Mg & 3.237 & 0.473 & 267.8& 3.307 &0.485	 &262.4	&	&		&	\\
$^{40}$Mg & 3.247 & 0.456 & 269.7& 3.320 &0.470	 &262.8	&	&		&	\\
\hline
\hline
\end{tabular}
\end{table}

\begin{table}
\caption{\label{tab:table4}Same as Table \ref{tab:table3}, for Al and Si 
isotopes.}
\begin{tabular}{|c|c|c|c|c|c|c|c|c|c|c|c|}
\hline
Nucl. & \multicolumn{3}{|c|}{RMF} & \multicolumn{3}{|c|}{SHF}     
&  \multicolumn{3}{|c|}{Exp.} \\
\cline{2-10}
          &$r_{ch}$&$\beta_2$&BE&$r_{ch}$&$\beta_2$&BE& $r_{ch}$&$\beta_2$&BE\\
\hline
$^{24}$Al&3.097	 &0.388	 &182.3	&3.174	 &0.413	 &185.0	&	&		&	183.6	\\
$^{25}$Al&3.072	 &0.381	 &197.7	&3.164	 &0.430	 &199.5	&	&		&	200.5	\\
$^{26}$Al&3.052	 &-0.275 &207.8	&3.122	 &0.315	 &211.4	&	&		&	211.9	\\
$^{27}$Al&3.053	 &-0.292 &221.9	&3.092	 &0.204	 &222.7	&3.061	&		&	225.0	\\
$^{28}$Al&3.037	 &-0.208 &238.6	&3.105	 &0.202	 &232.5	&	&		&	232.7	\\
$^{29}$Al&3.033	 &-0.141 &245.6	&3.126	 &0.241	 &241.5	&	&		&	242.1	\\
$^{30}$Al&3.070	 &-0.184 &253.8	&3.139	 &0.194	 &248.7	&	&		&	247.8	\\
$^{31}$Al&3.101	 &-0.205 &259.8	&3.161	 &-0.192 &256.0	&	&		&	255.0	\\
$^{32}$Al&3.103	 &-0.111 &261.2	&3.162	 &0.020	 &262.6	&	&		&	259.2	\\
$^{33}$Al&3.165	 &-0.333 &269.4	&3.183	 &0.000	 &269.8	&	&		&	264.7	\\
$^{34}$Al&3.134	 &-0.108 &275.1	&3.198	 &0.090	 &271.7	&	&		&	267.3	\\
$^{35}$Al&3.167	 &0.268	 &274.1	&3.229	 &0.250	 &274.4	&	&		&	272.5	\\
$^{36}$Al&3.173	 &-0.189 &277.7	&3.254	 &0.320	 &277.4	&	&		&	274.4	\\
$^{37}$Al&3.208	 &0.355	 &281.5	&3.278	 &0.371	 &280.1	&	&		&	278.6	\\
$^{38}$Al&3.214	 &-0.254 &283.5	&3.288	 &0.378	 &281.4	&	&		&	280.3	\\
$^{39}$Al&3.236	 &-0.299 &286.7	&3.383	 &-0.121 &287.1	&	&		&		\\
$^{40}$Al&3.257	 &-0.336 &290.4	&3.316	 &0.403	 &284.2	&	&		&		\\
$^{41}$Al&3.278	 &-0.370 &290.6	&3.338	 &-0.367 &285.9	&	&		&		\\
$^{42}$Al&3.281	 &-0.355 &291.2	&3.341	 &-0.339 &286.2	&	&		&		\\
$^{43}$Al&3.282	 &-0.338 &292.2	&3.341	 &-0.312 &286.6	&	&		&		\\
$^{44}$Al&3.274	 &-0.288 &293.6	&3.340	 &-0.282 &287.0	&	&		&		\\
$^{45}$Al&3.271	 &-0.263 &293.5	&3.338	 &-0.250 &287.6	&	&		&		\\
$^{46}$Al&3.359	 &0.341	 &294.5	&3.326	 &-0.129 &287.7	&	&		&		\\
$^{47}$Al&3.246	 &0.090	 &294.8	&3.318	 &-0.004 &288.7	&	&		&		\\
$^{48}$Al&3.319	 &-0.252 &294.0	&3.347	 &-0.060 &287.6	&	&		&			\\
$^{28}$Si&3.122	 &-0.331 &232.1	&3.190	 &-0.350 &233.6	&3.122	&0.407	&	236.5	\\
$^{29}$Si&3.035	 &0.001	 &240.7	&3.176	 &-0.272 &243.1	&3.118	&		&	245.0	\\
$^{30}$Si&3.070	 &0.148	 &250.6	&3.170	 &-0.210 &252.6	&3.134	&0.315	&	255.6	\\
$^{31}$Si&3.108	 &-0.180 &259.1	&3.182	 &-0.199 &261.7	&	&		&	262.2	\\
$^{32}$Si&3.137	 &-0.201 &268.5	&3.200	 &-0.200 &270.5	&	&0.217	&	271.4	\\
$^{33}$Si&3.131	 &-0.084 &275.6	&3.196	 &0.010	 &278.1	&	&		&	275.9	\\
$^{34}$Si&3.148	 &0.000	 &284.4	&3.220	 &0.000	 &286.3	&	&0.179	&	283.4	\\
$^{35}$Si&3.161	 &-0.083 &287.4	&3.226	 &0.010	 &289.5	&	&		&	285.9	\\
$^{36}$Si&3.186	 &-0.162 &291.5	&3.150	 &0.150	 &292.4	&	&0.259	&	292.0	\\
$^{37}$Si&3.200	 &0.238	 &295.4	&3.269	 &0.247	 &295.9	&	&		&	294.3	\\
$^{38}$Si&3.218	 &0.281	 &299.8	&3.290	 &0.310	 &298.2	&	&0.249	&	299.9	\\
$^{39}$Si&3.224	 &0.263	 &302.4	&3.298	 &0.292	 &301.4	&	&		&	301.5	\\
$^{40}$Si&3.272	 &-0.301 &306.0	&3.310	 &-0.280 &304.0	&	&		&	306.5	\\
$^{41}$Si&3.295	 &-0.336 &310.1	&3.349	 &-0.329 &307.2	&	&		&	307.9	\\
$^{42}$Si&3.318	 &-0.369 &314.6	&3.330	 &-0.350 &310.0	&	&		&		\\
$^{43}$Si&3.320	 &-0.356 &315.2	&3.377	 &-0.339 &311.1	&	&		&		\\
$^{44}$Si&3.322	 &-0.342 &316.2	&3.380	 &-0.300 &311.6	&	&		&		\\
$^{45}$Si&3.316	 &-0.308 &317.5	&3.374	 &-0.282 &312.9	&	&		&		\\
$^{46}$Si&3.303	 &-0.262 &319.3	&3.370	 &-0.240 &313.5	&	&		&		\\
$^{47}$Si&3.345	 &-0.298 &319.8	&3.340	 &0.030	 &314.3	&	&		&		\\
$^{48}$Si&3.263	 &0.001	 &321.8	&3.350	 &0.000	 &315.4	&	&		&		\\
$^{49}$Si&3.290	 &0.045	 &321.1	&	 & 	 &	&	&		&		\\
$^{50}$Si&3.341	 &-0.159 &321.5	&	 &	 &	&	&		&		\\
$^{51}$Si&3.358	 &-0.135 &321.2	&	 &	 &	&	&		&		\\
$^{52}$Si&3.371	 &0.082	 &321.4	&	 &	 &	&	&		&		\\
$^{53}$Si&3.391	 &0.042	 &321.6	&	 &	 &	&	&		&		\\
$^{54}$Si&3.415	 &0.000	 &322.3	&	 &	 &	&	&		&		\\
\hline
\hline
\end{tabular}
\end{table}

\begin{table}
\hspace{1.9cm}
\caption{\label{tab:table5} Same as Table \ref{tab:table3}, for P and S isotopes.
}
\begin{tabular}{|c|c|c|c|c|c|c|c|c|c|c|c|c|c|}
\hline
Nucl.  & \multicolumn{3}{|c|}{RMF} & \multicolumn{3}{|c|}{SHF}     
&  \multicolumn{3}{|c|}{Exp.} \\
\cline{2-10}
          &$r_{ch}$&$\beta_2$&BE&$r_{ch}$&$\beta_2$&BE& $r_{ch}$&$\beta_2$&BE\\
\hline
%
$^{30}$P&	3.138	&	0.130	&	246.3	&	3.189	&	0.026	&	249.9	&		&		&	250.6	\\
$^{31}$P&	3.158	&	0.205	&	258.3	&	3.201	&	0.105	&	261.1	&	3.189	&		&	262.9	\\
$^{32}$P&	3.174	&	-0.143  &	267.1	&	3.216	&	0.069	&	270.9	&		&		&	270.9	\\
$^{33}$P&	3.201	&	-0.183	&	277.5	&	3.246	&	-0.167	&	280.5	&		&		&	281.0	\\
$^{34}$P&	3.201	&	-0.082	&	285.8	&	3.248	&	0.001	&	289.9	&		&		&	287.2	\\
$^{35}$P&	3.216	&	-0.001	&	295.4	&	3.265	&	0.000	&	299.2	&		&		&	295.6	\\
$^{36}$P&	3.227	&	0.120	&	299.5	&	3.272	&	0.007	&	303.3	&		&		&	299.1	\\
$^{37}$P&	3.246	&	0.209	&	305.0	&	3.290	&	0.148	&	307.4	&		&		&	305.9	\\
$^{38}$P&	3.260	&	0.250	&	310.4	&	3.313	&	0.240	&	311.7	&		&		&	309.6	\\
$^{39}$P&	3.275	&	0.288	&	316.1	&	3.334	&	0.301	&	316.1	&		&		&	315.9	\\
$^{40}$P&	3.281	&	0.274	&	320.1	&	3.343	&	0.290	&	319.6	&		&		&	319.2	\\
$^{41}$P&	3.288	&	0.261	&	324.4	&	3.355	&	0.295	&	322.7	&		&		&	324.2	\\
$^{42}$P&	3.306	&	0.301	&	327.3	&	3.371	&	0.320	&	325.6	&		&		&	326.3	\\
$^{43}$P&	3.346	&	-0.323	&	331.7	&	3.398	&	-0.320	&	329.0	&		&		&	330.7	\\
$^{44}$P&	3.346	&	-0.302	&	333.3	&	3.398	&	-0.293	&	330.6	&		&		&		\\
$^{45}$P&	3.315	&	0.222	&	335.4	&	3.397	&	-0.264	&	332.4	&		&		&		\\
$^{46}$P&	3.342	&	-0.251	&	337.5	&	3.397	&	-0.237	&	334.2	&		&		&		\\
$^{47}$P&	3.341	&	-0.232	&	340.0	&	3.399	&	-0.218	&	336.0	&		&		&		\\
$^{48}$P&	3.381	&	-0.271	&	341.2	&	3.379	&	0.034	&	337.4	&		&		&		\\
$^{49}$ P&		3.328	&	0.088	&	343.2	&		3.387	&	0.012	&	339.3	&		&		&		\\
$^{50}$ P&		3.353	&	0.101	&	343.7	&		3.414	&	-0.061	&	339.2	&		&		&		\\
$^{51}$ P&		3.397	&	-0.166	&	344.7	&		3.437	&	0.068	&	339.4	&		&		&		\\
$^{52}$ P&		3.403	&	0.109	&	345.2	&		3.462	&	0.079	&	339.7	&		&		&		\\
$^{53}$ P&		3.428	&	0.109	&	346.3	&		3.487	&	0.089	&	340.1	&		&		&		\\
$^{54}$ P&		3.447	&	0.074	&	346.6	&		3.502	&	0.016	&	340.5	&		&		&		\\
$^{55}$ P&		3.468	&	0.037	&	347.4	&		3.525	&	0.001	&	341.2	&		&		&		\\

$^{33}$S&3.241	 &0.197	&275.5	&3.276	 &0.119	 &278.9	&	&		&280.4	\\
$^{34}$S&3.257	 &-0.168 &286.5	&3.300	 &-0.160 &289.3	&3.285	&0.252	&291.8	\\
$^{35}$S&3.260	 &-0.078 &295.7	&3.300	 &-0.006 &299.6	&	&		&298.8	\\
$^{36}$S&3.273	 &0.002	 &306.2	&3.310	 &0.000	 &309.6	&3.299	&0.168	&308.7	\\
$^{37}$S&3.285	 &0.152	 &311.6	&3.319	 &-0.008 &315.1	&	&		&313.0	\\
$^{38}$S&3.300	 &0.228	 &318.6	&3.340	 &0.210	 &320.2	&	&0.246	&321.1	\\
$^{39}$S&3.312	 &0.264	 &325.3	&3.354	 &0.248	 &326.5	&	&		&325.4	\\
$^{40}$S&3.325	 &0.299	 &332.4	&3.370	 &0.300	 &332.1	&	&0.284	&333.2	\\
$^{41}$S&3.331	 &0.287	 &337.7	&3.381	 &0.294	 &336.9	&	&		&337.4	\\
$^{42}$S&3.338	 &0.277	 &343.2	&3.390	 &0.290	 &341.0	&	&0.300	&344.1	\\
$^{43}$S&3.359	 &0.318	 &347.2	&3.413	 &0.326	 &344.7	&	&		&346.7	\\
$^{44}$S&3.381	 &0.367	 &351.0	&3.440	 &0.370	 &348.3	&	&0.254	&351.8	\\
$^{45}$S&3.375	 &0.312	 &353.4	&3.430	 &0.311	 &350.4	&	&		&354.7	\\
$^{46}$S&3.371	 &0.258	 &356.6	&3.420	 &0.250	 &352.5	&	&		&	\\
$^{47}$S&3.385	 &0.257	 &358.5	&3.428	 &-0.214 &354.8	&	&		&	\\
$^{48}$S&3.400	 &0.259	 &360.8	&3.430	 &-0.200 &356.6	&	&		&	\\
$^{49}$S&3.403	 &0.227	 &362.9	&3.430	 &0.127	 &358.8	&	&		&	\\
$^{50}$S&3.403	 &0.189	 &365.5	&3.440	 &0.120	 &360.8	&	&		&	\\
$^{51}$S&3.427	 &0.188	 &366.4	&3.459	 &-0.090 &361.8	&	&		&	\\
$^{52}$S&3.451	 &0.183	 &367.6	&3.490	 &-0.140 &362.5	&	&		&	\\
$^{53}$S&3.463	 &0.158	 &369.1	&3.508	 &-0.113 &363.6	&	&		&	\\
$^{54}$S&3.477	 &0.139	 &371.0	&3.530	 &0.000	 &364.7	&	&		&	\\
$^{55}$S&3.494	 &0.105	 &371.4	&3.541	 &0.030	 &365.4	&	&		&	\\
\hline
\hline
\end{tabular}
\end{table}

The theoretical prediction of drip nuclei are very important after 
the discovery of $^{40}$Mg and $^{42}$Al \cite{baumann07}. These two
nuclei are considered to be beyond the drip-line (neutron-unbound) in 
some of the mass calculations \cite{moller95,samyn04}. 
The discovery of these two isotopes, suggests the existence
of drip-line somewhere in the heavier side. Thus, the study
of these isotopes is beyond the scope of the existing mass models
\cite{moller95,samyn04}. In the present RMF/SHF calculations, the newly 
discovered $^{40}$Mg and $^{42}$Al are well within the drip-line.
Also, a point of caution, it may be possible that if we allow triaxial 
deformation in the calculation, then we may get one minimum as a saddle 
point and another one as triaxial 
minimum. But this calculation is out of scope in our paper, we are dealing
with axial deformed code by using NL3 and SkI4 parameter sets where mostly 
we find similar results in both the formalisms. This type of prescriptions 
are used in many of the earlier publications \cite{maharana92}.

\subsection{Neutron configuration}
Analyzing the neutron configuration for these exotic nuclei, we notice
that, for lighter isotopes
of Ne, Na, Mg, Al, Si, P and S the oscillator shell $N_{osc} = 3$ is empty
in the [$N_{osc}, n_3, \Lambda$]$\Omega^\pi$.
However, the $N_{osc} = 3$  shell gets occupied gradually with increase of
neutron number. In case of Na, $N_{osc} = 3$ starts filling up at $^{33}$Na 
with quadrupole moment deformation parameter $\beta_2 = 0.356$ and $-0.179$ 
with occupied orbits $[330]\frac{1}{2}^{-}$ and $[303]\frac{7}{2}^{-}$, 
respectively. 
The filling of $N_{osc} = 3$ goes on
increasing for Na with neutron number and it is $[330]\frac{1}{2}^{-}$,
$[310]\frac{1}{2}^{-}$, $[321]\frac{3}{2}^{-}$ and $[312]\frac{5}{2}^{-}$ 
at $\beta_2 = 0.472$ for $^{39}$Na.
Again for the oblate solution the occupation is  $[301]\frac{1}{2}^{-}$,
$[301]\frac{3}{2}^{-}$, $[303]\frac{5}{2}^{-}$ and $[303]\frac{7}{2}^{-}$ 
for $\beta_2 = -0.375$ for $^{39}$Na.
In case of Mg isotopes, even for $^{30,32}$Mg, the $N_{osc} = 3$ shell has
some occupation for the low-lying excited states near the Fermi surface. 
For $^{30}$Mg (at $\beta_2$ = 0.599 with BE = 237.7 MeV) the $N_{osc}$ = 3 
orbit is $[330]\frac{7}{2}^{-}$ and for $^{32}$Mg  is $[330]\frac{1}{2}^{-}$ 
(BE = 248.8 MeV at $\beta_2 = 0.471$). With the increase
of neutron number in Mg and Si isotopic chains, the oscillator
shell with $N_{osc} = 3$ gets occupied more and more. 

In Tables ($4-6$) the results for the ground state solutions are displayed.
Thus, the prolate solutions have more binding than the
oblate one for Ne, Na, Mg and S isotopes. In some cases, like
$^{24-30}$Ne the prolate and oblate solutions are in degenerate
states. For example, $^{24}$Ne has BE = 188.9 and 189.1 MeV at $\beta_2 = 0.278$
and $-0.259$ respectively. Contrary to this, the ground state solutions 
for Al and Si 
are mostly oblate. For example, $^{34}$Al has BE = 269.9 and 275.1 MeV
at $\beta_2 = 0.159$ and $-0.108$ respectively. In such cases, the prolate
solutions are in low-lying excited intrinsic state. 
Note that in many cases, there exist low laying superdeformed states.

It is important to list some of the limitations of the results due 
to the input parameters, mostly comes from $E_{pair}$ and $E_{cm}$ energies.
As one can see from Fig. \ref{deltaenergy}, in many cases there are solutions 
of different shapes lying only a few MeV higher, sometimes even degenerate 
with the ground states. Such a few MeV difference is within the uncertainty 
of the predicted binding energies. 
A slight change in the pairing parameter, among others,
may alter the prediction for the ground state shape. With few MeV uncertainty 
in ground state binding energies, by reassigning the ground state 
configurations, the deformation may change completely, and make the predictions 
close to each other and agree with the FRDM predictions as well. 

\subsection{Quadrupole deformation}
\begin{figure}
\includegraphics[width=1.0\columnwidth,clip=true]{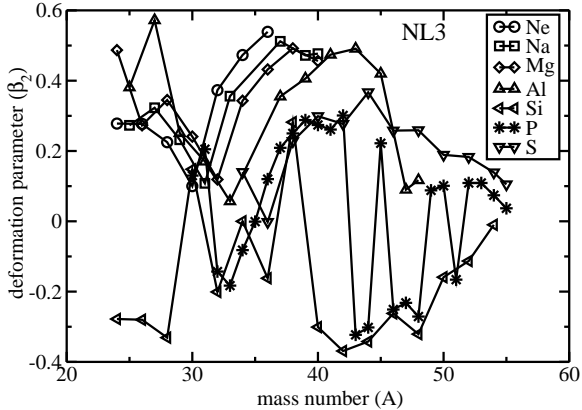}
\caption{\label{deformation}The ground state quadrupole deformation
parameter $\beta_2$ versus mass number A for Ne, Na, Mg, Al, Si, P 
and S isotopes near the drip-line with NL3 parameter set.
}
\end{figure}

The ground and low-lying excited state deformation systematics for some
of the representative nuclei for Ne, Na, Mg, Al, Si, P and S are analyzed.
In Fig. \ref{deformation}, the ground state quadrupole deformation parameter $\beta_2$
is shown as a function of mass number for Ne, Na, Mg, Al, Si, P and S.
The $\beta_2$ value goes on increasing with mass number for Ne, Na and Mg
isotopes near the drip-line.
The calculated quadrupole deformation parameter
$\beta_2$ for $^{34}$Mg is 0.59 which compares well with the recent
experimental measurement of Iwasaki et al \cite{iwasaki01}
($\beta_2 = 0.58\pm 0.06$). It found that this superdeformed state is 3.2 MeV
above than the ground band. Again, the magnitude of $\beta_2$ for
the drip nuclei reduces with neutron number N and again increases.
A region of maximum deformation is found for almost all the nuclei
as shown in the figure. It so happens in cases like, Ne, Na, Mg and
Al that the isotopes are maximum deformed which may be comparable to
superdeformed near the drip-line. For Al and Si isotopes,
in general, we find oblate solutions in the ground configurations 
(see Table $5$). In many of the cases, the low-lying superdeformed 
configuration are clearly visible and some of them can be seen in Fig. 
~\ref{deformation}.

\subsection{Shape coexistence}
One of the most interesting phenomena in nuclear structure physics is the
shape coexistence \cite{maharana92,ang93,sarazin00,egido04}. In some cases
of the nuclei, considered near the drip-line, the ground state
configuration accompanies a low-lying excited state. In a few cases, it so
happens that these two solutions are almost degenerate in energy. 
For example, in the RMF calculation, the ground state binding energy
of $^{24}$Ne is 189.1 MeV with $\beta_2 = -0.259$ and the binding energy of
the excited low-lying configuration at $\beta_2 = 0.278$ is 188.9 MeV.
The difference
in BE of these two solutions is only 0.179 MeV. Similarly the
solution of prolate-oblate binding energy difference in SkI4 is 0.186 MeV
for $^{30}$Mg with $\beta_2 = -0.183$ and 0.202. This type of degenerate 
solutions are observed in most of the isotopes near the drip-line.
It is worthy to mention that in the truncation of the basis space, an 
uncertainty of $\leq$ 1 MeV in total binding energy may occur. 
However, this uncertainty in convergence does not effect to determine 
the shape co-existence, because both the solutions are 
obtained by using the same model space of $N_F=N_B=12$.

\begin{figure}
\includegraphics[scale=0.35]{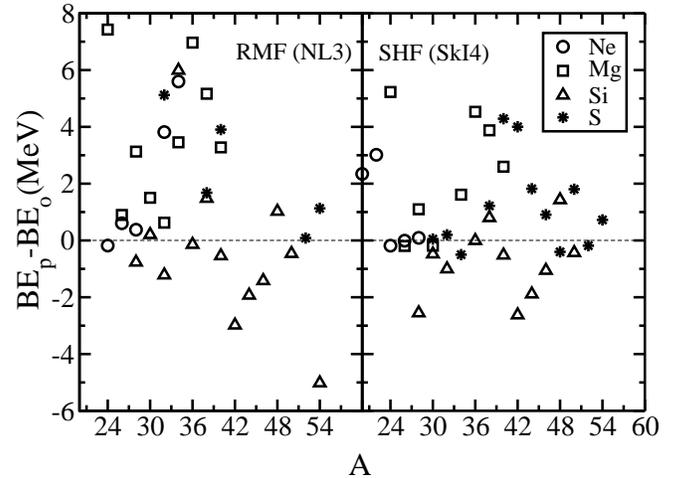}
\caption{\label{deltaenergy}{ The difference in binding energy
between the prolate-oblate solutions is shown for even-even Ne, Mg, Si and S
isotopes near the neutron drip-line with NL3 and SkI4 parameter sets.
}}
\end{figure}

To show in a quantitative way, we have plotted the prolate-oblate
binding energy difference ($BE_p-BE_o$) in Fig. \ref{deltaenergy}. 
The left hand side of the figure
is for relativistic and the right side is the nonrelativistic results.
From the figure, it is clear that an island of shape coexistence
isotopes are available for Mg and Si isotopes. These shape coexistence
solutions are predicted taking into account the intrinsic binding energy.
However, the actual quantitative energy difference of ground and excited
configuration can be given by performing configuration mixing (mixing such 
as in the generator coordinate method(GCM) \cite{onishi66}) after the 
angular momentum projection \cite{ang93}.

\subsection{Two neutron separation energy ($S_{2n}$) }
\begin{figure}
\includegraphics[scale=0.35]{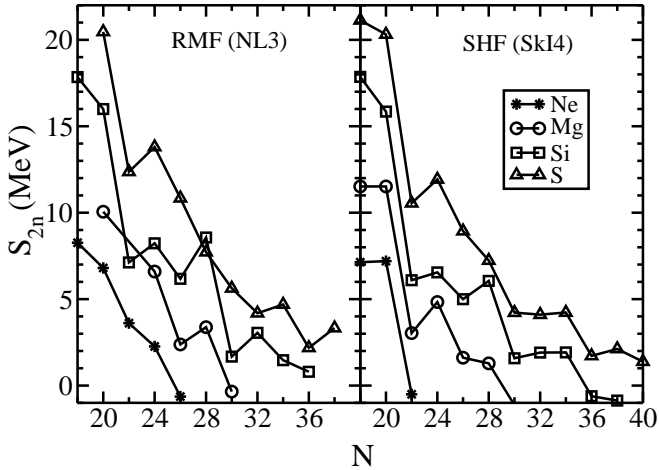}
\caption{\label{s2n}{ The two-neutron separation $S_{2n}$ energy
versus neutron number N for neutron-rich Ne, Mg, Si and S isotopes.
}}
\end{figure}

The appearance of new and the disappearance of known
magic number near the neutron drip-line is a well discussed topic
currently in nuclear structure physics \cite{ozawa00,jha03}. Some of the
calculations in recent past predicted the disappearance of the known magic
number N = 28 for the drip-line isotopes of Mg and S 
\cite{werner94,ren01,hamamoto07}.
However, magic number 20 retains its magic properties even for the drip-line
region.
In one of our earlier publications, \cite{patra} we analyzed the 
spherical shell gap
at N = 28 in $^{44}$S and its neighboring $^{40}$Mg and $^{42}$Si using NL-SH
\cite{sharma93} and TM2 parameter sets \cite{toki94}. The spherical shell
gap at N = 28 in $^{44}$S was found to be intact for the TM2 and is broken for
NL-SH parametrization.
Here, we plot the two-neutron separation energy $S_{2n}$ for Ne,
Mg, Si and S for the even-even nuclei near the drip-line (Fig. \ref{s2n}). 
The known magic number N = 28 is noticed to be absent in $^{44}$S. 
On the other hand, the appearance of a sudden decrease in $S_{2n}$ energy at 
N = 34 in SHF result is quite prominent, which is not clearly visible in RMF 
prediction. This is just two units ahead than the experimental shell closure 
at N = 32 \cite{kanungo02}.

\subsection{ Superdeformation and Low $\Omega$ parity doublets}
\begin{figure}
\includegraphics[scale=0.35]{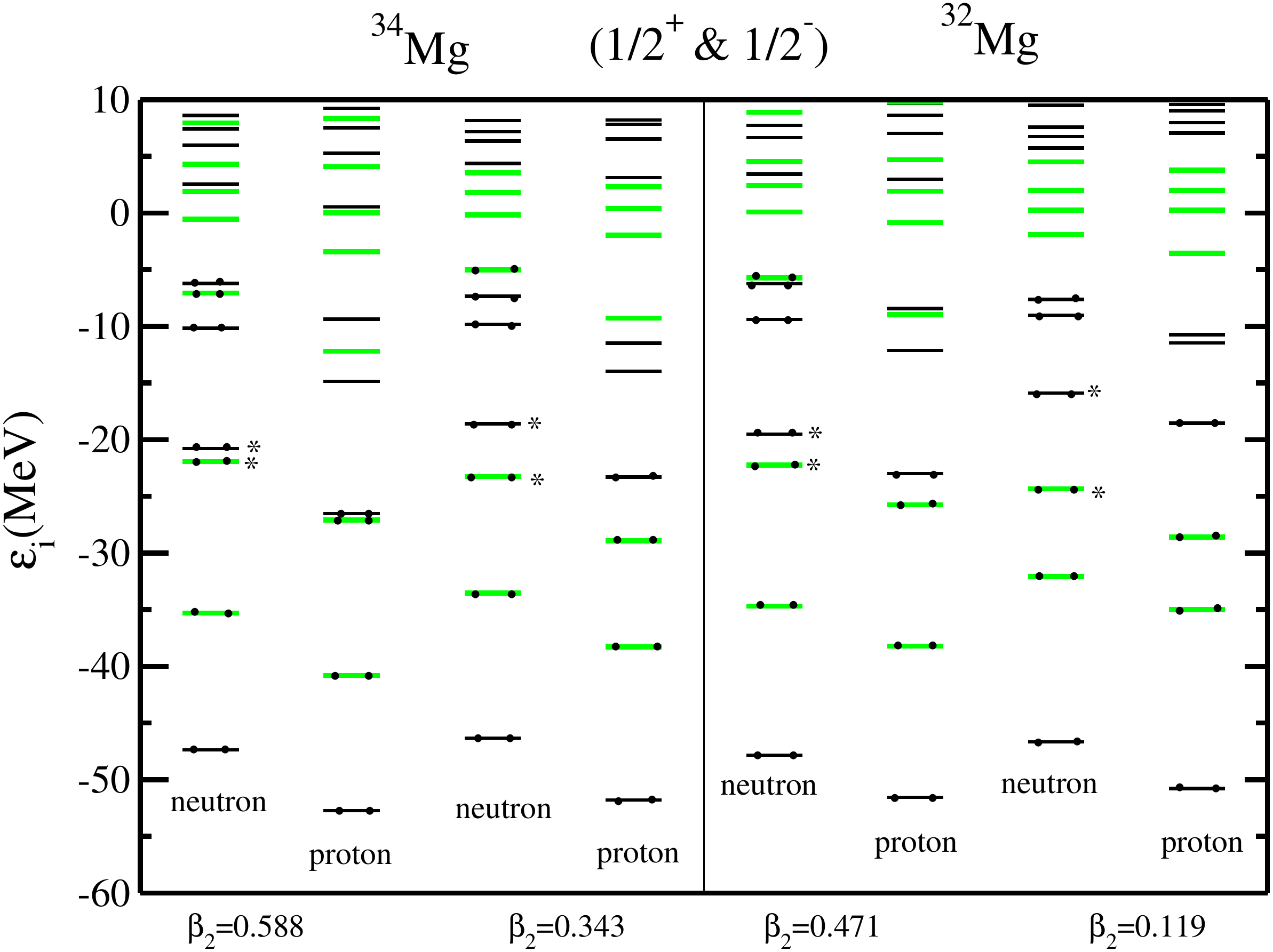}
\caption{\label{mg}
{ The $\frac{1}{2}^+$ and $\frac{1}{2}^-$ intrinsic single-particle 
states for
the normal and superdeformed state for $^{32}$Mg and $^{34}$Mg.
Doublets are noticed for the SD intrinsic states only. The 
${\pm \frac{1}{2}^-}$ states are denoted by green lines and the 
${\pm \frac{1}{2}^+}$ states are denoted by black.}
}
\end{figure}

\begin{figure}
\includegraphics[scale=0.35]{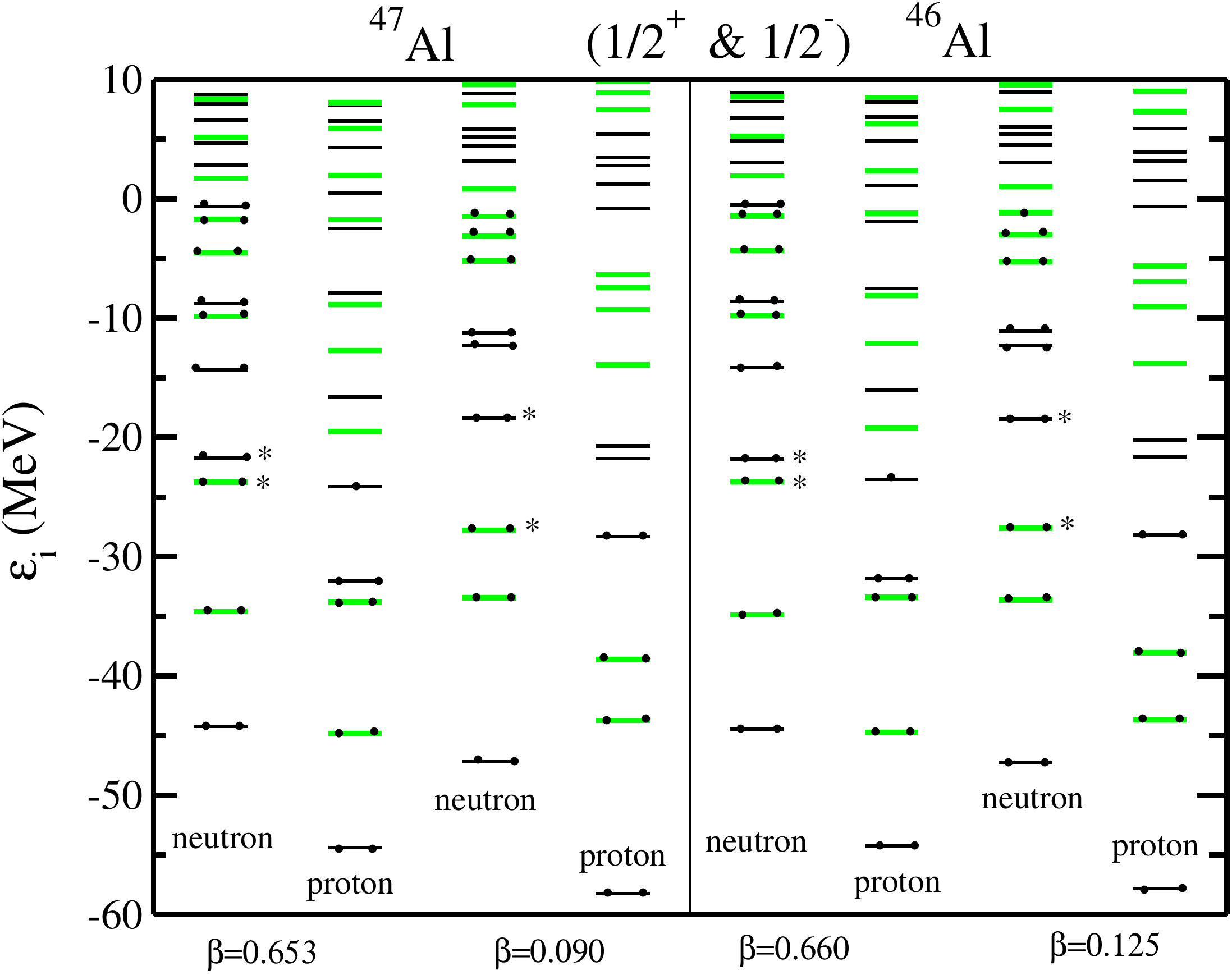}
\caption{\label{al}{ Same as Fig. \ref{mg} for $^{46}$Al
and $^{47}$Al.
}}
\end{figure}

The deformation-driving $m = \frac{1}{2}^-$orbits come down in energy
in superdeformed solutions from the shell above, in contrast to
the normal deformed solutions. The occurrence of approximate
$\frac{1}{2}^+$, $\frac{1}{2}^-$ parity doublets (degeneracy of $|m|^{\pi}$=
$\frac{1}{2}^+$, $\frac{1}{2}^-$ states) for the superdeformed solutions 
are clearly
seen in Figs. \ref{mg} and \ref{al} where excited superdeformed configurations
for $^{32}$Mg, $^{34}$Mg and for $^{46}$Al, $^{47}$Al are given (RMF solutions).
For each nucleus, we have compared the normal deformed
$(\beta_{2} \sim 0.1-0.3)$ and the superdeformed configurations and
analyzed the deformed orbits.
The $\frac{1}{2}^+$ and $\frac{1}{2}^-$ states for the single particle 
levels are
shown in Fig. \ref{mg} (for $^{32}$Mg and $^{34}$Mg) and Fig. \ref{al} for 
$^{47}$Al and $^{46}$Al. The occupation of neutron states (denoted by $m^\pi$)
in $^{47}$Al and $^{46}$Al is given in Table \ref{parity}. 
In both $^{47}$Al and $^{46}$Al two neutrons occupying oblate driving $f_{\frac{7}{2}}$
$m=\frac{7}{2}$ orbits in normal deformation are unoccupied in the superdeformed (SD) 
case; instead two neutrons occupy the very prolate deformation driving [440]1/2
orbits (raising $n_{\frac{1}{2}^+}$ to 10) which is a superposition of $g_{\frac{9}{2},\frac{7}{2}}$ 
$d_{\frac{5}{2},\frac{3}{2}}$ $s_{\frac{1}{2}}$ orbits of $N_{osc}=4$ origin. In $^{46}$Al one  
$m=\frac{3}{2}^-$ neutron
shift to $m=\frac{1}{2}^-$, enhancing the prolate deformation. 
It is to be  emphasized that the 
deformations of occupied orbits of self-consistent SD solutions are more 
(than their normal deformed counterparts) because of mixing among the shells.

\begin{table}
\caption{\label{parity}Occupation of neutron orbits $m^{\pi}$ in $^{47}$Al
and $^{46}$Al driving by the deformation.}
\begin{tabular}{|c|c|ccccccccc|}
\hline
A	& $\beta_2$ & $n_{\frac{1}{2}^+}$ & $n_{\frac{1}{2}^-}$ & 
$n_{\frac{3}{2}^+}$ & $n_{\frac{3}{2}^-}$ & $n_{\frac{5}{2}^+}$
& $n_{\frac{5}{2}^-}$ & $n_{\frac{7}{2}^+}$ & $n_{\frac{7}{2}^-}$
& $n_{\frac{9}{2}^+}$ \\
\hline
$^{47}$Al & 0.09 & 8 & 10 & 4 & 6 & 2 & 2 & 0 & 2 & 0 \\
$^{47}$Al & 0.672 & 10 & 10 & 4 & 6 & 2 & 2 & 0 & 0 & 0  \\
\hline
$^{46}$Al & 0.109& 8 & 9 & 4 & 6 & 2 & 2 & 0 & 2 & 0 \\ 
$^{46}$Al & 0.701& 10 & 10 & 4 & 5 & 2 & 2 & 0 & 0 & 0 \\
\hline
\end{tabular}
\end{table}

\subsubsection{Structure of Superdeformed Configuration:}

We discus some clear and important characteristics of superdeformed solutions
 ( $\beta \sim 0.5 $ or more) obtained in mean field models as compared to the 
normal solutions of smaller deformation. Since the lowering and occupation 
of the deformation-driving $\Omega=\frac{1}{2}$ orbits from the shell above the usual
valence space is so important in producing superdeformation we have emphasized
their role in this discussion. There is the occurrence of $\frac{1}{2}^+$, 
$\frac{1}{2}^-$ orbits 
close together in energy (doublets) below and near the Fermi surface of the 
self-consistent superdeformed solutions. This feature also occurs broadly in 
Nilsson orbits at asymptotically large prolate deformations (see the Nilsson
diagrams in Bohr and Mottelson vol. II  \cite{BohM}).


\subsubsection{Some features of superdeformed solutions:} 

In normal deformed case, the deformed orbits of a major shell form a 
``band''-like set of orbits, distinctly separated from the major shell above 
and below  (see Fig. \ref{al} for $^{47}$Al ($\beta$=0.09) and $^{46}$Al 
($\beta$ = 0.125)).
Thus physical states obtained from such intrinsic states of low deformation
will be well separated in energy from those intrinsic states where 
excitation occurs across a major shell (a single nucleon excitation across a 
major shell means a change in parity and significant energy change for small 
deformation). 

The above mentioned ``band''-like separation of orbits of major shells of 
unique parity is quite lost in the case of superdeformation (see Fig. \ref{al},
$\beta$=0.653 of $^{47}$Al and $\beta$=0.660 of $^{46}$Al).
The ``band''-like orbits now spread in energy (both downward and upward) and 
orbits of successive major shells come closer to each other in energy; an 
inter-mingling  of orbits of different parities (see Figs. \ref{mg}, \ref{al}).
This is a significant structural change from the case of small deformation.
This has also been seen in the case of $^{84}$Zr in Hartree-Fock study
\cite{crp07,crp99}.

We would like to emphasize that in the self-consistent models 
(Skyrme-HF and RMF) the deformation of the nucleus is the result of the 
deformation of the self-consistently occupied individual orbits:
\begin{equation}
Q=\sum_{i (occupied)}  {q_i \cdots}
\end{equation}
The occupation of the more deformation driving orbits from the shell above the 
valence space and the unoccupation of oblate driving orbits (e.g. $f_{\frac{7}{2}}$,
$m=\pm\frac{7}{2}$) contribute much to configuration mixing and the lowering of 
$m=\frac{1}{2}$ orbits and to generation of the quadrupole deformation. 
Because of coming together in energy of $m=\frac{1}{2}^+$ and $\frac{1}{2}^-$ 
orbits , it is easy to see that superdeformed intrinsic states of two 
different parities 
for a particular K quantum number can be formed which will be close to each 
other in energy. This will lead to parity doublets in band structures.
For the neutron-rich nuclei being discussed here, the protons
are quite well bound and possible low energy excitations will be those of 
neutrons near the Fermi surface.
\begin{equation}
|\phi_K> = |\phi^p_{K_p}>|\phi^n_{K_n}> \cdots, 
\end{equation}
where $K_p$ and $K_n$ are the K quantum numbers for proton and neutron 
configurations (K=$K_p$ + $K_n$).

\subsubsection{Examples of parity doublet configurations:}

\begin{figure}
\includegraphics[scale=0.35]{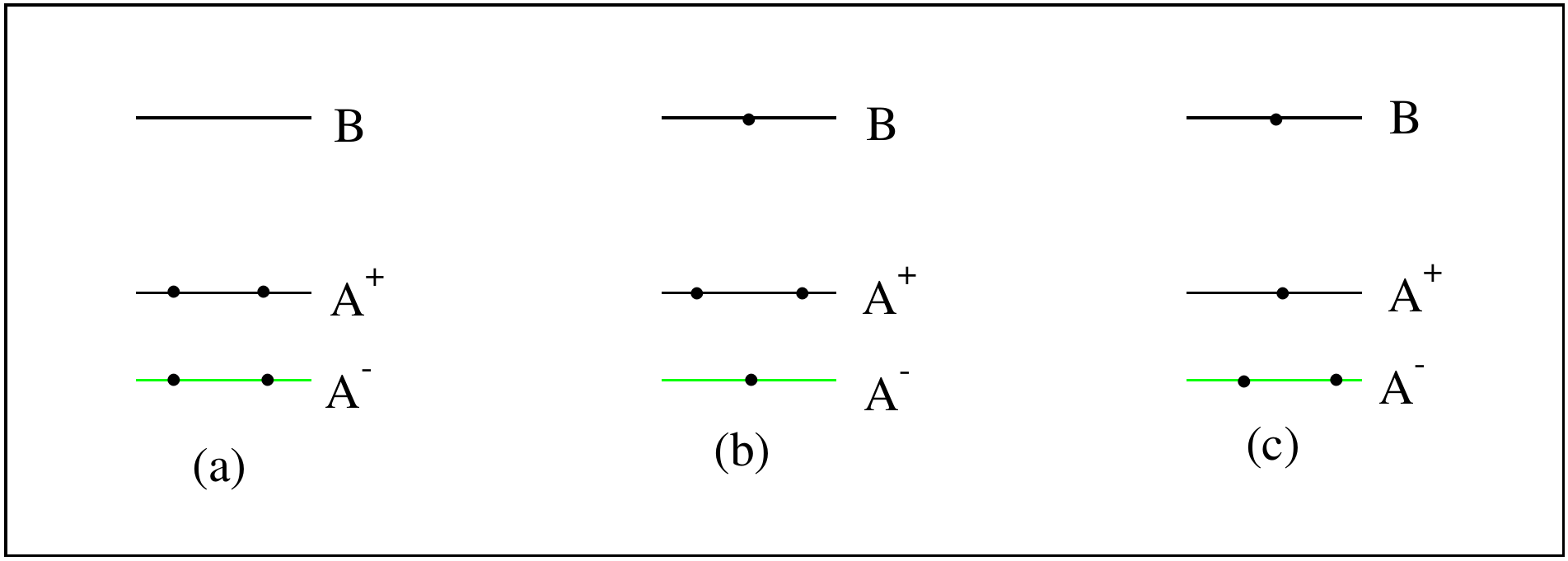}
\caption{\label{cartoon}{ 
With parity doublet of occupied orbits $A^+$, $A^-$ (having m= $\pm$ 1/2 
and +ve and -ve parities) and an unoccupied orbits B possible occupation of 
neutrons are shown in configurations (a), (b) and (c). The two excited 
configurations (b) and (c) have the same $K_n$ value and represent two 
excited bands of different parities (parity doublet). Such situation
can occur for neutron configurations in superdeformed $^{47}$Al and 
$^{32}$Mg, $^{34}$Mg (Figs. \ref{al}, \ref{mg}).
}}
\end{figure}

We illustrate schematically possible parity doublet of configurations 
for neutrons in Fig. \ref{cartoon}, the proton configuration 
$|\phi^p_{K_p}>$ being fixed. We show here the last few neutron occupations
of superdeformed solutions and rearrangements near the Fermi surface.
In Fig. \ref{cartoon}, (b) and (c) are parity doublet of configurations.
$A^+ \rightarrow A^- $ transition between (b) and (c) configurations 
is of odd parity multipole nature. 

Thus, in summary, we find a systematic
behaviour of the low $\Omega$ (particularly $\frac{1}{2}^+$ and 
$\frac{1}{2}^-$) prolate
deformed orbits for the superdeformed solutions. 
We notice (from the plot of the orbits) that there is occurrence 
of $\frac{1}{2}^+$ and $\frac{1}{2}^-$
orbits very close to each other in energy for the superdeformed (SD) shape.
Such $\frac{1}{2}^+$, $\frac{1}{2}^-$ degenerate orbits  occur not only for the
well-bound orbits but also for the unbound states. 
For example, the doublet of neutron orbits $[220]\frac{1}{2}^+$ and 
$[101]\frac{1}{2}^-$ 
are $4$ MeV apart from each other in the normal deformed prolate 
solutions; but they become degenerate in the superdeformed (SD) solutions 
(shown by * in Figs. \ref{mg} and \ref{al} for Mg, Al).
More such doublets are easily identified
(Figs. \ref{mg} and \ref{al}) for superdeformed solutions of $^{32,34}$Mg and $^{46,47}$Al.
In fact it is noticed that the $\Omega = \frac{1}{2}$ states of unique parity, 
seen clearly well separated in energy from the usual parity orbits
in the normal deformed solutions, occurs closer to them in energy 
for the SD states, showing a degenerate parity doublet structure. 
In fact, for SD solution the $\frac{1}{2}^+$ and $\frac{1}{2}^-$ orbits are 
intermixed in 
the energy plot; while for the normal deformation they occur in 
distinct groups. This is true both in the Skyrme Hartree Fock and the RMF 
calculations. 

This can be seen by examining the $\frac{1}{2}^+$ and 
$\frac{1}{2}^-$ orbits for small and large deformations in Fig. \ref{mg}.
This can lead
to parity mixing and octupole deformed shapes for the SD structures
\cite{crp07}. Parity doublets and octupole deformation for superdeformed
solutions have been discussed for $^{84}Zr$ \cite{crp07,crp99}. 
There is much interest for the experimental study of the spectra of
neutron-rich nuclei in this mass region \cite{miller}. The highly deformed
structures for the neutron-rich Ne-Na-Mg-Al nuclei are interesting
and signature of such superdeformed configurations (with parity doublet
structure) should be looked for.

\section{Summary and Conclusions}
In summary, we calculate the ground and low-lying excited state properties,
like binding energy and quadrupole deformation $\beta_2$ using RMF(NL3)
formalism for Ne, Na, Mg, Si, P and S isotopes, near the neutron drip-line region.
In general, we find large deformed solutions for the neutron-drip
nuclei which agree well with the experimental measurement. 
The calculation is also repeated in the frame-work of nonrelativistic 
Hartree-Fock formalism with Skyrme interaction SkI4. Both the relativistic 
and non-relativistic 
results are comparable to each other for the considered mass region. 
In the present calculations a large number of low-lying intrinsic 
superdeformed excited states are predicted in many of the isotopes and 
some of them are reported.
From binding energy point of view, i.e. the sudden fall in $S_{2n}$ value,
the breaking of N = 28 magic number and the likely appearance of a new magic 
number at N = 34 noticed in our non-relativistic calculations, in contrast
with RMF finding. This is an indication of more binding than the neighbouring
isotopes. However to confirm N = 34 as a magic /non-magic number 
more calculations are needed. A deformed nucleus has a collective low 
lying $2^+$ state. Also, a spherical nucleus can have a fairly low lying 
collective $2^+$ state (e.g. Sn nuclei) because of quadrupole collectivity.
In this study we find that, for the SD shape, the low $\Omega$ orbits 
(particularly $\Omega = \frac{1}{2}$)
become more bound and nearly degenerate with the orbits of opposite 
parity, i.e. they show a parity doublet structure.
Closely lying parity-doublet band structures and enhanced odd parity 
multipole transitions are possible for the superdeformed shapes. 

CRP was supported during this work by Project SR/S2/HEP-37/2008 of Department 
of Science and Technology, Govt. of India.


\begin{thebibliography}{99}
\bibitem{navin00} A. Navin et al., Phys. Rev. Lett. {\bf 85}, 266 (2000);
        	  H. Iwasaki et al., Phys. Lett. B {\bf 491}, 8 (2000);
        	  H. Iwasaki et al., Phys. Lett. B {\bf 481}, 7 (2000).
\bibitem{motobayashi95} T. Motobayashi et al., 
	Phys. Lett. B {\bf 346}, 9 (1995).
\bibitem{tanaka10} K. Tanaka et al.,
	Phys. Rev. Lett. {\bf 104}, 062701 (2010)
\bibitem{iwasaki01} H. Iwasaki et al.,
	Phys. Lett. B {\bf 522}, 227 (2001).
\bibitem{baumann07} T. Baumann et al.,
	Nature {\bf 449}, 1022 (2007).
\bibitem{moller95} P. M\"oller, R. J. Nix, W. D. Myers and W. J. Swiatecki, 
        At. Data and Nucl. Data Tables {\bf 59}, 185 (1995).
\bibitem{ozawa00} A. Ozawa, T. Kobayashi, T. Suzuki, K. Yoshida and 
        I. Tanihata, 
	Phys. Rev. Lett. {\bf 84}, 5493 (2000).
\bibitem{mina} 
	T. Minamisono, T. Ohtsubo, I. Minami, S. Fukuda, A. Kitagawa, 
	M. Fukuda, K. Matsuta, Y. Nojiri, S. Takeda, H. Sagawa 
	and H. Kitagawa, 
	Phys. Rev. Lett. {\bf 69}, 2058 (1992).
\bibitem{tan96}  I. Tanihata, 
	J. Phys. G {\bf 22}, 157 (1996).
\bibitem{nakamura11} T. Nakamura, 
	J. Phys. Conf. Ser. {\bf 312}, 082006 (2011);
        T. Nakamura et al., Phys. Rev. Lett. {\bf 103}, 262501 (2009).
\bibitem{vautherin72} D. Vautherin and D. M. Brink,
	Phys. Rev. C {\bf 5}, 626 (1972).
\bibitem{reinhard89} P. -G. Reinhard, 
	Rep. Prog. Phys. {\bf 52}, 439 (1989).
\bibitem{cha97} E. Chabanat, P. Bonche, P. Hansel, J. Meyer and R. Schaeffer,
	Nucl. Phys. A {\bf 627}, 710 (1997).
\bibitem{bender03} M. Bender, Paul-Henri Heenen and P. -G. Reinhard,
	Rev. Mod. Phys. {\bf 75}, 121 (2003).
\bibitem{lunney03} D. Lunney, J. M. Pearson and C. Thibault,
	Rev. Mod. Phys. {\bf 75}, 1021 (2003).
\bibitem{stone03} J. R. Stone, J. C. Miller, R. Koncewicz, P. D. Stevenson and 
	M. R. Strayer,
	Phys. Rev. C {\bf 68}, 034324 (2003).
\bibitem{stone07} 
	J. R. Stone and P. -G. Reinhard,
	Prog. Part. Nucl. Phys. {\bf 58}, 587 (2007).
\bibitem{erler11} J. Erler, P. Kl\"upfel and P. -G. Reinhard,
	J. Phys. G: Nucl. Part. Phys. {\bf 38}, 033101 (2011).
\bibitem{takashi12} T. Nakatsukasa,
	Prog. Theor. Exp. Phys. 01A207 (2012).
\bibitem{sero86} B. D. Serot and J. D. Walecka, 
	Adv. Nucl. Phys. {\bf 16}, 1 (1986).
\bibitem{ring90} Y. K. Gambhir, P. Ring and A. Thimet, 
	Ann. Phys. (N.Y.) {\bf 198}, 132 (1990).
\bibitem{vretenar05} D. Vretenar, A. V. Afanasjev, G. A. Lalazissis and P. Ring,
	Physics Reports {\bf 409}, 101 (2005).
\bibitem{meng06} J. Meng et al., 
	Prog. Part. Nucl. Phys. {\bf 57}, 470 (2006).
\bibitem{niksic11} T. Nik\^si\'c, D. Vretenar and P. Ring,
	Prog. Part. Nucl. Phys. {\bf 66}, 519 (2011). 
\bibitem{paar07} N. Paar, D. Vretenar, E. Khan and Gianluca Col\'o,
	Rep. Prog. Phys. {\bf 70}, 691 (2007).
\bibitem{rei95} P. -G. Reinhard and H. F. Flocard,
	Nucl. Phys. A {\bf 584}, 467 (1995).
\bibitem{cha98} E. Chabanat, P. Bonche, P. Haensel, J. Meyer and R. Schaeffer,
        Nucl. Phys. A {\bf 635}, 231 (1998).
\bibitem{pres82} M. A. Preston and R. K. Bhaduri, 
	{\it Structure of Nucleus, Addison-Wesley Publishing Company}, 
	Ch. 8, page 309 (1982).
\bibitem{patra93} S. K. Patra, 
	Phys. Rev. C {\bf 48}, 1449 (1993).
\bibitem{dech80} J. Decharg and D. Gogny, 
	Phys. Rev. C {\bf 21}, 1568 (1980).
\bibitem{madland} D. G. Madland and J. R. Nix,
	Nucl. Phys. A {\bf 476}, 1 (1981).
\bibitem{bcs-delta} S. J. Krieger, P. Bonche, H. Flocard, P. Quentin and 
	 M. S. Weiss, 
	Nucl. Phys. A {\bf 517}, 275 (1990).
\bibitem{delta}H. Zhang et al., 
	Eur. Phys. J. A {\bf 30}, 519 (2006).
\bibitem{raman} S. Raman, C. W. Jr. Nestor and P. Tikkanen,
	At. Data and Nucl.  Data Tables {\bf 78}, 1 (2001).
\bibitem{audi12}M. Wang, G. Audi, A. H. Wapstra, F. G. Kondev, M. MacCormick, 
	X. Xu and B. Pfeiffer, 
	Chin. Phys. C {\bf 36}, 1603 (2012).
\bibitem{angeli13} I. Angeli, K. P. Marinova,
	At. Data and Nucl. Data Tables {\bf 99}, 69 (2013). 
\bibitem{ripka} G. Ripka, 
	Adv. Nucl. Phys. {\bf 1}, 183 (1968);
	W. H. Bassichis, B. Giraud and G. Ripka,
	Phys. Rev. Lett. {\bf 15}, 980 (1965).
\bibitem{mott69} A. Bohr and B. R. Mottelson,
	Nuclear Structure Vol. I W A Benjamin, Inc (1969).
\bibitem{lala99} G. A. Lalazissis, S. Raman and P. Ring, 
	At. Data and Nucl.  Data Tables {\bf 71}, 1 (1999).
\bibitem{patra01} S. K. Patra, M. Del. Estel, M. Centelles and X. Vi\~nas, 
	Phys. Rev. C {\bf 63}, 024311 (2001).
\bibitem{patra91} S. K. Patra and C. R. Praharaj,
	Phys. Rev. C {\bf 44}, 2552 (1991).
\bibitem{doba84} J. Dobaczewski, H. Flocard and J. Treiner, 
	Nucl. Phys. A {\bf 422}, 103 (1984).
\bibitem{meng96}J. Meng and P. Ring,
	Phys. Rev. Lett. {\bf 77}, 3963 (1996).
\bibitem{del01} M. Del Estal, M. Centelles, X. Vi\~nas and S. K. Patra,
	Phys. Rev. C {\bf 63}, 044321 (2001).
\bibitem{stoi98} M. Stoitsov, P. Ring, D. Vretenar and G. A. Lalazissis,
	Phys. Rev. C {\bf 58}, 2086 (1998).
\bibitem{nakada02} H. Nakada and M. Sato,
	Nucl. Phys. A {\bf 699}, 511 (2002).
\bibitem{shan03} Shan-Gui Zhou, J. Meng and P. Ring,
	Phys. Rev. C {\bf 68}, 034323 (2003).
\bibitem{shan10} Shan-Gui Zhou, J. Meng, P. Ring and En-Guang Zhao,
	Phys. Rev. C {\bf 82}, 011301(R) (2010).
\bibitem{patra05} P. Arumugam, B. K. Sharma, S. K. Patra and R. K. Gupta,
	Phys. Rev. C {\bf 71}, 064308 (2005).
\bibitem{hfb} J. Meng and P. Ring,
	Phy. Rev. Lett. {\bf 77}, 3963 (1996).
\bibitem{elliott55} J. P. Elliott and T. H. R. Skyrme,
	Proc. R. Soc. London A {\bf 232}, 561 (1955).
\bibitem{negele70} J. W. Negele,
	Phys. Rev. C {\bf 1}, 1260 (1970).
\bibitem{toki94} Y. Sugahara and H. Toki, 
	Nucl. Phys. A {\bf 579}, 557 (1994). 
\bibitem{lala97} G. A. Lalazissis, K. K\"onig and P. Ring,
	Phys. Rev. C {\bf 55}, 540 (1997).
\bibitem{brown98} B. A. Brown, 
	Phys. Rev. C {\bf 58}, 220 (1998).
\bibitem{sn} R. C. Nayak and L. Satpathy,
	At. Data and Nucl. Data Tables {\bf 98}, 616 (2012). 
\bibitem{frdm} P. M\"oller, J. R. Nix and K. -L. Kratz, 
	At. Data and Nucl. Data Tables {\bf 66}, 131 (1997).
\bibitem{samyn04} M. Samyn, S. Goriely, M. Bender and J. M. Pearson, 
	Phys. Rev. C {\bf 70}, 044309 (2004).
\bibitem{maharana92} J. P. Maharana, Y. K. Gambhir, J. A. Sheikh and P. Ring,
        Phys. Rev. C {\bf 46}, R1163 (1992).
\bibitem{ang93} S. K. Patra and C. R. Praharaj,
	Phys. Rev. C {\bf 47}, 2978 (1993).
\bibitem{sarazin00} F. Sarazin et al.,
	Phys. Rev. Lett. {\bf 84}, 5062 (2000).
\bibitem{egido04} J. L. Egido, L. M. Robledo, R. R. Rodriguez-Guzman, 
	Phys. Rev. Lett. {\bf 93}, 282502 (2004). 
\bibitem{onishi66} N. Onishi and S. Yoshida, 
	Nucl. Phy. {\bf 80}, 367 (1966). 
\bibitem{jha03} T. K. Jha, M. S. Mehta, S. K. Patra, B. K. Raj and R. K. Gupta,
	PRAMANA -J. Phys. {\bf 61}, 517 (2003);
       L. Satpathy and S. K. Patra, Nucl. Phys.  A {\bf 722}, 24c (2003);
      R. K. Gupta, M. Balasubramaniam, Sushil Kumar, S. K. Patra,
      G. M\"unzenberg and W. Greiner, J. Phys. G {\bf 32}, 565 (2006);
      R. K. Gupta, S. K. Patra and W. Greiner, 
	Mod. Phys. Lett. A {\bf 12}, 1327 (1997).
\bibitem{werner94} T. R. Werner, J. A. Sheikh, W. Nazarewicz,
         M. R. Strayer, A. S. Umar and M. Misu, 
	Phys. Lett. B {\bf 335}, 259 (1994).
\bibitem{ren01} Ren Zhongzhou, Z. Y. Zhub, Y. H. Cai and Xu Gongou, 
	Phys. Lett. B {\bf 380}, 241 (1994).
\bibitem{hamamoto07} I. Hamamoto, 
	Phys. Rev. C {\bf 76}, 054319 (2007);
	{\it ibid} C {\bf 85}, 064329 (2012).
\bibitem{patra} R. K. Gupta, S. K. Patra and W. Greiner, 
	Mod. Phys. Lett. A {\bf 12}, 1317 (1997).
\bibitem{sharma93} M. M. Sharma, M. A. Nagarajan and P. Ring, 
	Phys. Lett. B {\bf 312}, 377 (1993).
\bibitem{kanungo02} R. Kanungo, I. Tanihata and A. Ozawa, 
	Phys. Lett. B {\bf 528}, 58 (2002).
\bibitem{BohM} A. Bohr and B. R. Mottelson, 
	Nuclear Structure, vol. II (, Chapter 5
\bibitem{crp07} C. R. Praharaj, {\it INT Workshop on "Nuclear Many-Body 
	Approaches for the $21^{st}$ Century"}, 
	September 24 - November 30, 2007,
	Institute for Nuclear Theory website,
	University of Washington, Seattle. 
\bibitem{crp99} C. R. Praharaj, {\it Structure of Atomic Nuclei},
        Ch. 4, page 108, Edited by Satpathy L, Narosa Publishers (Delhi) (1999).
\bibitem{miller} D. Miller et al., 
	Phys. Rev. C {\bf 79}, 054306 (2009).
\end{thebibliography}
\end{document}